\documentclass[aps,prl,twocolumn]{revtex4-1}

\usepackage{amssymb,color}
\usepackage{graphicx}
\usepackage{amsmath}
\usepackage{hyperref}

\begin{document}

\title{Anyonic statistics
revealed \\ by 
the Hong-Ou-Mandel dip for fractional excitations}

\author{T. Jonckheere, J. Rech, B. Gr\'emaud, T. Martin}

\affiliation{Aix Marseille Univ, Universit\'e de Toulon, CNRS, CPT, Marseille, France}

\date{\today}

\begin{abstract}
The fractional quantum Hall effect (FQHE) is known to host anyons, quasiparticles whose statistics is intermediate between bosonic and fermionic.
We show here that Hong-Ou-Mandel (HOM) interferences between excitations created 
by narrow voltage pulses on the edge states of a FQHE system at low temperature 
show a direct signature of anyonic statistics. The width
of the HOM dip is universally fixed by the thermal time scale,
independently of the intrinsic width of the excited fractional wavepackets.
This universal width can be related to the anyonic braiding of the incoming excitations with thermal
fluctuations created at the quantum point contact. We show
that this effect could be realistically observed with periodic trains of narrow voltage pulses using current
experimental techniques. 
\end{abstract}

\maketitle

Fractional quantum Hall effect (FQHE) is an important example of a many-body system  where electronic correlations have an essential impact.\cite{stormer99} When
a fraction $\nu$ of the states of the lowest Landau level is occupied, the system reaches a state which cannot be understood without electronic interactions. The well-known Laughlin wavefunction describes the highly correlated ground state of the FQHE when $\nu=1/(2n+1)$ for $n$ integer. The fundamental excitations of the FQHE are anyons: quasiparticles which bear a fractional charge, and obey fractional statistics.\cite{Arovas1984,stern08}
In a given Laughlin state, when two anyons are exchanged,  the system acquires a phase $\mbox{exp}(i \pi \nu)$, to be contrasted with the $\pm 1$ of bosonic/fermionic statistics.
More complex fractions exist, potentially hosting non-Abelian anyons, relevant for quantum computing applications.\cite{nayak08}

A fractional charge $e/3$ was experimentally observed for the Laughlin state  with $\nu=1/3$ more than twenty years ago, by measuring the shot noise across a quantum point contact (QPC)  in the tunneling regime, where individual fractional quasiparticles can tunnel between opposite edge states.\cite{saminadayar97,depicciotto97,Kapfer2019,Bisognin2019} 
Fractional statistics, however, has proved more difficult to observe. Only very recently, two different experiments have been able  to clearly show  specific signatures directly associated with the fractional statistics of anyonic quasiparticles.\cite{bartolomei20,nakamura20}

Electronic transport in FQHE occurs only through chiral edge modes, traveling along the boundary.  These can be used as 1d electron beams enabling to realize transport experiments inspired from quantum optics, such as the Hong-Ou-Mandel (HOM) interference experiment, where two identical photons are sent with a controlled time delay on a beam-splitter.\cite{hong87} The electronic counterpart was performed a few years ago in the integer QHE, where current correlations were shown to give precious information on the
electronic wavepackets and the many-body electronic state.\cite{bocquillon13,bocquillon14,jonckheere12,wahl14} Recently, two-particle time-domain intereferences were obtained in the FQHE, demonstrating that quasiparticles keep their coherence allowing for time-domain interference.\cite{taktak22}

In this work, we show that using narrow periodic pulses of voltage, periodically exciting fractional charges, and measuring the HOM noise at the output of a QPC, one obtains a signal which is directly related to the anyonic statistics. To this aim, we first explain the unique properties of the time-dependent tunneling current at a QPC when a single fractional quasiparticle is incident, which are associated with braiding of the fractional quasiparticle with the thermal anyonic excitations occurring at the QPC. Our quantitative predictions, obtained with perturbative calculations performed using the non-equilibrium Keldysh Green function formalism, could be checked with current experimental techniques, providing a relatively easy path for the study of fractional statistics. 

\begin{figure}
\includegraphics[width=8.5cm]{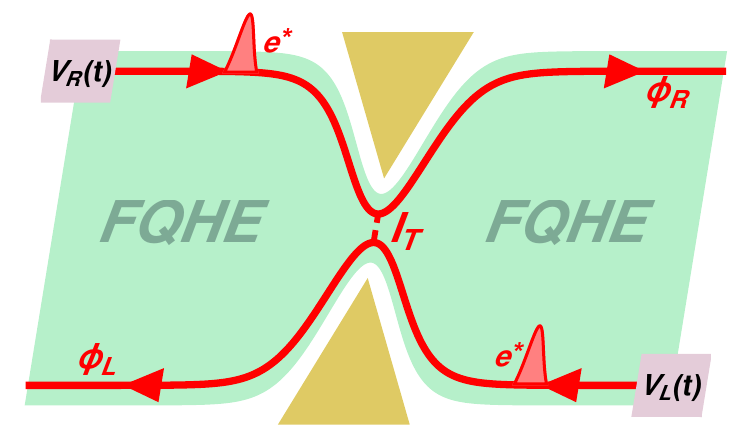}
\caption{The setup: a Hall bar in the Laughlin series, whose edge states are described by the bosonic fields $\phi_R$ and $\phi_L$, with a QPC at position $x=0$. The right- and left-moving edges are driven respectively by the time-dependent potential $V_R (t)$ and $V_L(t)$, resulting in a tunneling current $I_T$ in between edge.}
\label{fig:setup}
\end{figure}

We consider a FQH bar, with Laughlin filling factor $\nu=1/(2n+1)$ for $n$ integer, and describe the edge states in terms of the bosonic Hamiltonian
$H_0= \frac{v_F}{4 \pi} \int \!\! dx \sum_{\mu=R,L} (\partial_x \phi_{\mu})^2$
 where $\phi_{R/L}$  are the
bosonic fields describing the right-/left-moving edge states propagating with velocity $v_F$.\cite{wen95} 
A bosonization identity, $\psi_{R/L}(x) = U_{R/L}/(2 \pi a) e^{\pm i k_F x} e^{-i \sqrt{\nu} \phi_{R/L}(x)}$, relates the quasiparticle (QP) operator to the
bosonic field, with a a small cutoff parameter $a$ and $U_{R/L}$ a Klein factor.
The presence of a QPC (at $x=0$), in the weak backscattering regime,
 allows the tunneling of individual
QP of charge $e^*=\nu e$ between the two edges. This is described by the tunneling Hamiltonian
$H_T = \Gamma \psi_R^{\dagger}(0) \psi_L(0) + \mbox{H.c.}$ See
Fig.~\ref{fig:setup} for a sketch of the setup.

To better understand the importance of the anyonic statistics for tunneling at the QPC, 
let us first consider the somewhat simpler situation where
 a single QP of charge $e^*$ is incoming
on the $R$ edge. To this aim, when computing physical quantities (current, etc.),
we replace the ground state by a prepared state
$|\varphi \rangle = \psi_R^{\dagger}(-x_0,-\mathcal{T}) |0 \rangle$
where a single QP has been added at an initial time $-\mathcal{T}<0$.
Without loss of generality, we choose $x_0 = v_F \mathcal{T}$, such that the QP
reaches the QPC at $t=0$. We now proceed with the perturbative calculation
of the mean tunneling current $\langle I_T (t)\rangle$ at the QPC, using standard Keldysh Green
function formalism. The tunneling current operator is given by 
$I_T(t)= i e^* (\Gamma \psi_R^{\dagger}(0,t) \psi_L(0,t) - \mbox{H.c.})$.
To lowest order in $\Gamma$, the mean current is given by~\footnote{See Supplemental Material for further details.}
\begin{align}
\left \langle I_T(t) \right \rangle &= - \frac{i}{2} \int \!\! dt'
\sum_{\eta,\eta'}  \eta' 
\left\langle \varphi \left| T_K \;  I_T \left(  t^\eta \right) H_T \left( t'^{\eta'} \right) \right|\varphi \right\rangle ,
 \label{eq:IKeldysh}
\end{align}
where $T_K$ is time-ordering along the Keldysh contour, and $\eta,\eta'=\pm$ are Keldysh indices.
Using the bosonized form of the quasiparticle operators, and keeping in mind that
$x_0 = v_F \mathcal{T}$, we have
\begin{align}
\left \langle I_T(t) \right \rangle = \Gamma^2  \frac{e^*}{2} \int \!\! dt' &
\sum_{\eta,\eta'}  \eta'  \left[\mathcal{G} \left( \sigma^{\eta \eta'}_{t t'} (t-t')\right)\right]^2  
\nonumber \\
&  \times  \left[ \frac{\mathcal{G}(-t') \mathcal{G}(t)}{\mathcal{G}(t') \mathcal{G}(-t)}  -   \frac{\mathcal{G}(t') \mathcal{G}(-t)}{\mathcal{G}(-t') \mathcal{G}(t)}  \right] ,
\label{eq:ITint}   
\end{align}
with
\begin{equation}
\mathcal{G}(t)=
\frac{1}{2 \pi a} \left[ \frac{\mbox{sinh}(i \pi a /(\beta v_F))}
{\mbox{sinh}(i \pi a /(\beta v_F) - \pi t/\beta)}\right]^{\nu} ,
\end{equation} 
where $ \mathcal{G} ( \sigma^{\eta \eta'}_{t t'} (t-t'))=$ $\mathcal{G} ( 0;t^{\eta},t'^{\eta'})$ $=\langle 0 | T_K \psi^{\dagger} (0,t^{\eta}) \psi (0,t'^{\eta'}) | 0  \rangle$, with  $\sigma^{\eta \eta'}_{t t'}= \mbox{sign}(t-t') (\eta+\eta')/2 + (\eta'-\eta)/2$ accounting for the effect of time-ordering 
along the Keldysh contour. 
$\mathcal{G}$ is the quasiparticle Green function (identical for right and left movers),
directly obtained from its bosonic counterpart,
  with $\beta$ the inverse temperature. Note that the power $\nu$ leads to a slow decay of this Green function
   at long times since $\nu < 1$, up to the thermal time scale 
    $\tau_\text{Th} = \hbar \beta$.
  In the limit of vanishing cutoff
 $a \to 0$, it is easy to check that 
 $\mathcal{G}(t) / \mathcal{G}(-t)= \mbox{exp}(-\mbox{sign}(t) \times i \pi \nu)$. This directly arises from the nontrivial exchange properties of anyonic quasiparticles, exploiting their linear dispersion along the edge.\cite{Note1} It follows that the last factor of Eq.~\eqref{eq:ITint} can be simplified as
 \begin{equation}
 \frac{\mathcal{G}(-t') \mathcal{G}(t)}{\mathcal{G}(t') \mathcal{G}(-t)}   = \mbox{exp}\left(-i \,  \nu \int_{t'}^{t} \!\!\! d\tau \; 2 \pi \delta(\tau)\right) .
\end{equation} 
The current can thus be written as
\begin{align}
\left \langle I_T(t) \right \rangle = 2 i  e^* \Gamma^2 
&\int_{-\infty}^{t} \!\! dt' \; 
\sin\left(2 \pi \nu \int_{t'}^{t} \!\!\! d\tau \; \delta(\tau)\right) \nonumber \\
  & \times \left[ \mathcal{G}(t-t')^2 - \mathcal{G}(t'-t)^2\right]  .
\label{eq:Iqp}
\end{align}
 
One readily sees from Eq.~\eqref{eq:Iqp} that the tunneling current has remarkable properties, which are unique
 to fractional charge tunneling in the FQHE.\cite{Note1}
 It is of course zero for $t<0$, i.e. before the arrival
 of the $e^*$ QP.  
 On the other hand, for $t>0$, the $t'$ integration is restricted to the negative portion of the real axis,
  and the current is simply proportional to  $\sin(2 \pi \nu)$. This, in turn leads to a non-zero current even for a time $t$ taken long after the $e^*$ QP 
   has reached the QPC position,
   as a consequence of the slow decay in time of the Green function $\mathcal{G}(t-t')$. 
 
The mean current thus remains finite for a long time interval, set by the thermal time scale $\tau_\text{Th}$. We emphasize that this is in sharp contrast with the case of an incoming electron charge, since even for fractional edge states, the mean tunneling current is non-zero only at the specific time that the electron arrives at the QPC.~\cite{Note1}

This nontrivial behavior of the tunneling current after the arrival of a single QP of charge $e^*$ can
 be directly linked to the anyonic statistics of the fractional excitations.
 The phase $2 \pi \nu$ occurring for $t'<0<t$ can be understood qualitatively
from Eq.~\eqref{eq:IKeldysh} by considering the time-ordering of the right-moving edge operators  
($\psi_R$, $\psi_R^{\dagger}$). From the expressions of $| \varphi \rangle$, $I_T$ and $H_T$, one readily sees that the average current in Eq.~\eqref{eq:IKeldysh} involves a contribution of the form $T \psi_R(0) \psi_R^ {\dagger}(t) \psi_R(t') \psi_R^{\dagger}(0)$, as the prepared state ensures that the QP reaches the QPC at time 0.
For $t>0$ and $t'<0$, the time ordering thus requires to bring both $ \psi_R(0)$ and $\psi_R^{\dagger}(0)$
between the operators at $t$ and $t'$, yielding twice a phase  $\pi \nu$. On the opposite, if $t$ and $t'$ have the same sign, one can easily see that the exchanges needed for the ordering now contribute with opposite phases, thus giving a zero net result. 
 An equivalent point of view, developed in Refs.~[\onlinecite{lee19,lee20,morel22}] is
 to see the expression of Eq.~\eqref{eq:IKeldysh} as the interference between a process where
 a quasiparticle/quasihole (QP/QH) excitation
  is created at the location of the QPC at time $t'$, before the passage of the  $e^*$ QP, and
  another where the QP/QH is created at time $t$, after the passage of the $e^*$ QP. 
  In both points of view, the tunneling current is non-zero because of the braiding of
  the incoming fractional QP with a thermal QP/QH excitation created at the QPC.
  This braiding results from the anyonic statistics, giving a non-trivial 
  phase $\pi \nu$ when two quasiparticles are exchanged.

We now show that the same tunneling current, with the same signature of fractional statistics, can be obtained by applying a short voltage pulse which excites a fractional average charge. This is a highly nontrivial statement, as it is known that such a voltage pulse does not create the same many-body state as the one obtained by adding 
a single quasiparticle on top of the ground state.\cite{rech17} 

The presence of an external time-dependent voltage bias leads to an extra term in the total Hamiltonian, of the form
 $H_{V}= -\frac{2 e \sqrt{\nu}}{v_F} V(x,t) \partial_x \phi_R$.
 The voltage can be taken into account by using the following transformation:
$\phi(x,t) = \phi^{(0)}(x,t) + e \sqrt{\nu} \int_{-\infty}^{t} \!\!\! dt' \; V(x',t')$ ,
with $x' = x -v_F (t-t')$ and where $\phi^{(0)}(x,t)$ is the equilibrium bosonic field.\cite{rech17}
Assuming that the voltage is applied on a long contact, we can simplify
$\int_{-\infty}^{t} \!\!\! dt' V(v_F(t'-t),t') \simeq \int_{-\infty}^{t} \!\!\! dt' V(t')$.
This leads to a time-dependent tunneling amplitude at the QPC
$\Gamma(t) = \Gamma \mbox{exp}[i e^* \int_{-\infty}^{t} dt' V(t')]$. 
Proceeding with the perturbative calculation of the tunneling current, one gets~\cite{Note1}
\begin{align}
\left \langle I_T (t) \right \rangle =
2 i e^* \Gamma^2 
&\int_{-\infty}^{t} \!\!\!\! dt' \; \sin \left( e^*\int_{t'}^{t} dt'' V(t'') \right) \nonumber \\
& \times \left[ \mathcal{G}(t-t')^2 - \mathcal{G}(t'-t)^2\right] .
\label{eq:Istart}
\end{align}
One can thus readily recover the result of Eq.~\eqref{eq:Iqp}, provided that one chooses a voltage pulse $V(t)= \frac{2 \pi}{e} \delta(t)$, which excites a mean charge $e^*=\nu e$. The tunneling current $\langle I_T (t) \rangle$ is the same for a single QP of charge 
$e^*$ arriving on the QPC, or when applying a very short voltage pulse $V(t)$ exciting a mean charge $e^*$. 

\begin{figure}
 \includegraphics[width=0.48\textwidth]{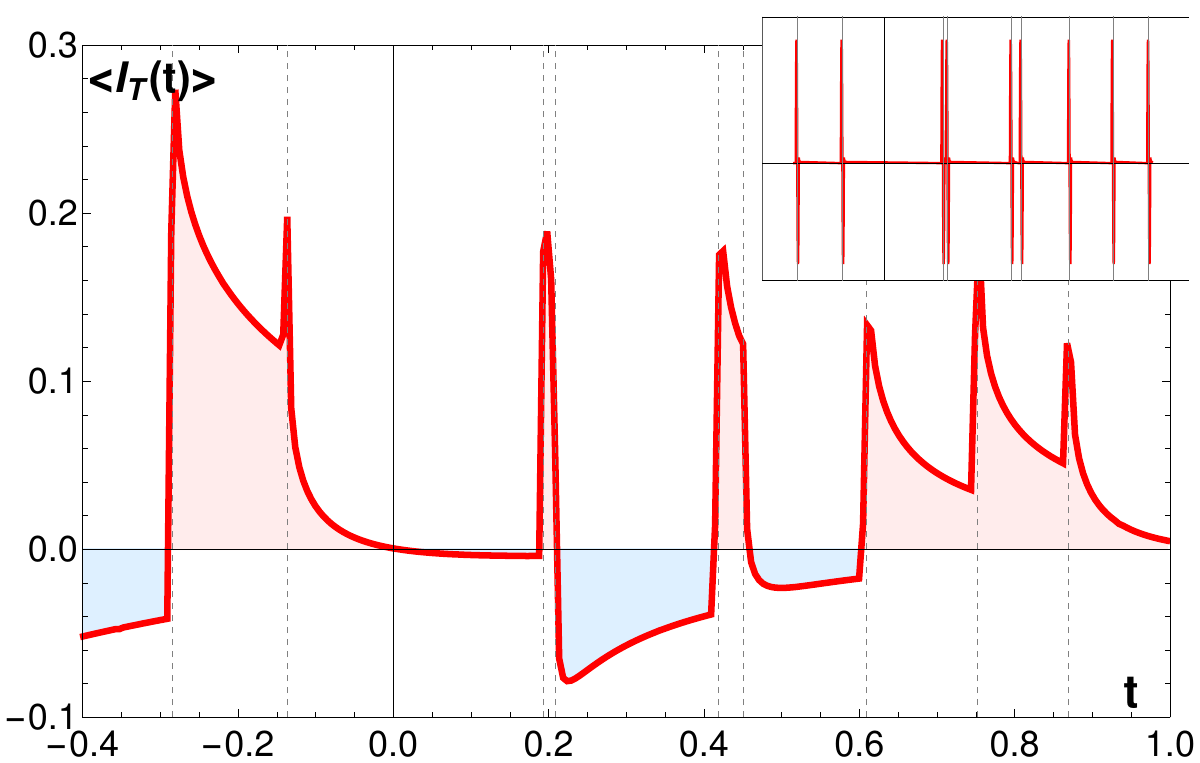}
 \caption{Mean current $\left \langle I_T (t) \right \rangle$ (in units of $e/\beta$)
  as a function of $t$ (in units of $\beta$) corresponding
 to Eq.~\eqref{eq:Istart} with $\nu=1/3$, 
 for a random ensemble of short pulses of width $\beta/100$, each carrying a charge $e/3$. The
  arrival times at the QPC are shown as dashed
 vertical lines (pulses for $t<-0.4 \beta$ are not shown).
 Inset: same figure for pulses carrying a charge $e$.}
 \label{fig:Ioftrandom}
 \end{figure}

This picture is further generalized by considering a voltage $V(t)$ composed of several short pulses of charge $e^*$. There, the phase
of the sine term counts the number of fractional charges $e^*$ that have passed through the QPC,
each of them contributing a phase $2 \pi \nu$. This then has important consequences for the tunneling current. For example, at filling factor $\nu =1/3$, when
two short fractionally charged pulses arrive at the QPC with a time delay much smaller than the thermal scale,
the main contribution to the current in Eq.~\eqref{eq:Istart} comes with a factor
$\sin(4 \pi/3) <0$, making it negative. 
 Fig.~\ref{fig:Ioftrandom} shows the current for an ensemble of short pulses 
 (each with a charge $e^*$, and a width $\delta t \ll \beta$).
  The dashed lines show the arrival times of the pulses at the QPC. We
 see that the current decreases slowly in absolute value after each pulse reaches the QPC,
 reflecting the slow decrease of the QP Green function $\mathcal{G}$. More interestingly,
 the value of the current depends on the history of the pulses applied at earlier times.
 In particular, as argued above, the current can be negative when two pulses
  arrive at closely separated times (e.g. for $t$ between 0.2 and $0.4 \beta$).
  The inset of Fig.~\ref{fig:Ioftrandom} shows the equivalent picture when similar pulses,
  but carrying a charge $e$ rather than $e^*$, are incident on the QPC. 
There, the current is non-zero only when the pulse reaches the QPC, with no effect from earlier pulses. Note that the same expression for the tunneling current, Eq.~\eqref{eq:Istart}, allows one to describe a random stream of pulses, recovering known results for the collision between two Poissonian streams of charges e*.\cite{rosenow16}

 While the use of voltage pulses is routinely performed, the measurement of time-dependent currents still constitutes an experimental challenge in quantum Hall junctions. We now propose a simpler alternative, within grasp of modern experiments, in order to reveal the effect of anyonic statistics. This relies on the measurement of the HOM noise, i.e. the current correlations resulting from two individual voltage pulses of fractional charge colliding at the QPC
 with a controllable time delay.
 
 Let us first consider two narrow pulses of charge $e^*$,
 incoming on the two inputs of the QPC.
 The tunneling current noise is defined as
 \begin{equation}
 S(t,t') =  \left \langle T_K \; \delta I_{T}(t^{-}) \, \delta I_{T}(t'^{+}) \right \rangle,
  \end{equation}
 with $\delta I_{T} (t) = I_{T}(t) - \left \langle I_{T} (t) \right \rangle$,
 and $\pm$ are Keldysh indices.
  The HOM noise is the zero-frequency tunneling noise, when two pulses are
  incident on the QPC with a given time delay $\delta t$. It serves as a measure of the interference between the colliding excitations at the QPC.
  It can be written as~\cite{Note1}
\begin{align}
S_{HOM}(\delta t) =\frac{1}{2 S_{HBT}}  & \int_{-\infty}^{\infty} \!\!\!\! dt dt' 
   \; \mathcal{G}(t'-t)^2 \nonumber \\ 
& \times 
\left\{ \cos \left[2 \pi \nu f_{\delta t}\left(t,t'\right) \right] -1 \right\} ,
\label{eq:HOMsingle}
\end{align}
where $f_{\delta t}(t,t')$ is 1 if only one of the times $t$ or $t'$ is in the interval 
$[-\delta t/2,\delta t/2]$, and 0 otherwise, and normalization is given by  twice the value of the Hanbury-Brown Twiss (HBT) noise $S_{HBT}$~\cite{bocquillon12}.

 A very good approximation of Eq.~\eqref{eq:HOMsingle} (exact for $\delta t/\beta \to 0$), is given by 
 \begin{equation}
 S_{HOM}(\delta t) \overset{\delta t \ll \tau_\text{Th}}{\longrightarrow} 1- \exp \left(-2 \pi \nu \frac{|\delta t|}{\beta} \right) .
 \label{eq:HOMsingleResult}
 \end{equation}
This result shows a behavior typical of a HOM dip for long and short time delays. For very large $|\delta t|$, it saturates to 1 as
the two incident charges $e^*$ reach the QPC at very distant times without interfering, therefore reproducing twice the amount of the HBT noise.
For $\delta t =0$, the HOM dip drops all the way to 0, as a result of perfect interference between the 
two identical incoming charges. 
 This can be understood as a fractional charge injected from the left and
another one injected from the right braiding with opposite phases with the thermal excitations of
the QPC. At $\delta t=0$ these phases cancel exactly.
The most important result, however, is the behavior at intermediate $\delta t$:
Eq.~\eqref{eq:HOMsingleResult} shows that the width of the HOM dip is $\sim \beta$, set by the thermal time scale $\tau_\text{Th}$, independently
of the width of the incoming pulses. 
This is in sharp contrast with the conventional HOM dip, for
example between electronic wavepackets in the integer QHE~\cite{jonckheere12,Marguerite2016},
 where the dip width is directly proportional to that of the incoming wavepacket. 
 This striking result can be understood from our discussion of the tunneling current above. Indeed, we showed that, as a consequence of anyonic statistics and the braiding with thermal excitations, a single charge $e^*$ reaching the QPC creates a nonzero current up to times $\sim \beta$ after the tunneling event occurred.

Two charges incident on both inputs of the QPC thus interfere up to times set by the thermal time scale, which explains the width of the HOM
dip. The observation of such an HOM dip can thus provide a direct proof of the anyonic statistics of the incoming fractional charges.
We now show how a realistic periodic voltage bias with frequency $\omega$,
 sending pulses of charge $q \, e$ (with non-integer $q$), can
 be used to observe the HOM dip of width $\sim \beta$. For illustrative purposes,
 we consider a periodic voltage $V(t)$ consisting of Lorentzian pulses, also known as levitons,~\cite{lee93,keeling06,dubois13,dubois13b}
 but the results are independent of the actual shape of the voltage potential, as long as the pulse
 width is small compared to $\beta$.
 We use the Floquet formalism, where the essential ingredients are the coefficients $p_l$, which 
 correspond to the Fourier coefficients of the phase $\phi(t) = e^* \int_{-\infty}^t dt' V_{AC}(t')$
 created by the AC part of the time-dependent voltage $V(t)$.
 The DC part of the voltage leads to a mean charge $q e$ injected per period,
 with $q=e^* V_{DC}/ \omega$.
 We consider that the voltages $V_{R}(t)$ and $V_{L}(t)$, applied on the right and
 left edge respectively, differ by a time-shift $\delta t$ only, so that
  \begin{equation}
   V_L(t) = V_R(t-\delta t)= \frac{V_{DC}}{\pi} \sum_k \frac{\eta}{\eta^2+(t/T_0-k)^2} ,
  \end{equation} 
where $T_0=2 \pi/\omega$ is the period of the drive, and $\eta$ is the finesse.
 
\begin{figure}
\includegraphics[width=0.48\textwidth]{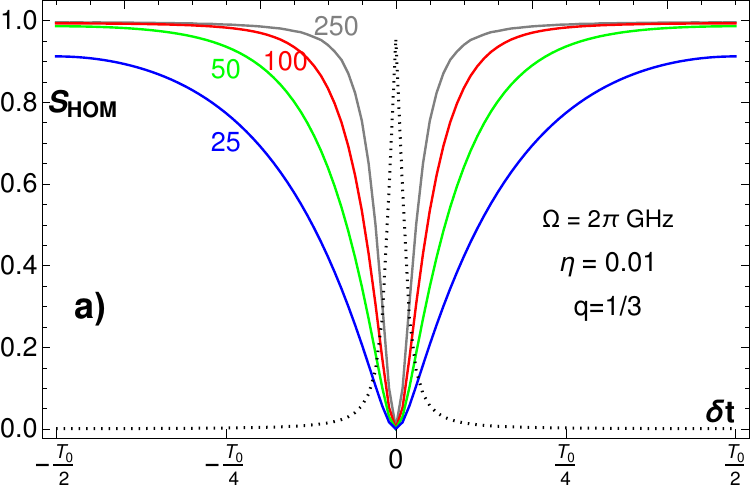}
\includegraphics[width=0.48\textwidth]{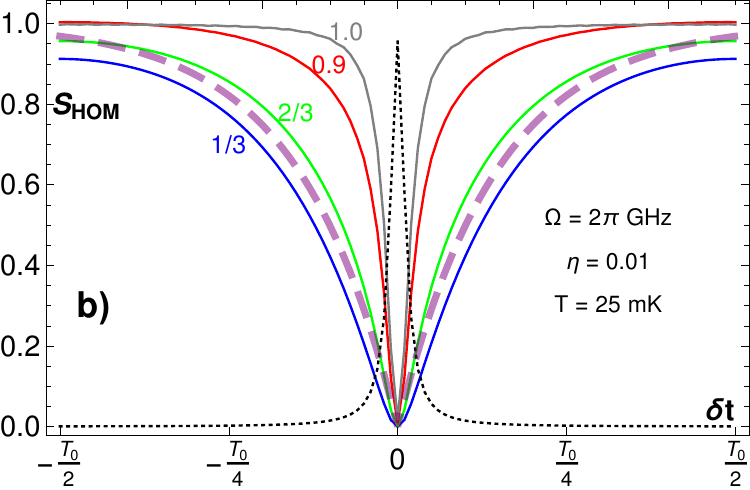}
\caption{HOM noise as a function of $\delta t$ 
for a filling factor $\nu=1/3$,
for $V(t)$ made of voltage pulses with Lorentzian shape of finesse $\eta=0.01$,
 with $\omega = 2 \pi/T_0= 2\pi$GHz. a) pulses of charge $e/3$,
 and $T$ in mK indicated near each curve. b) $T= 25$mK,
  and the charge of each pulse (in units of $e$) shown near each curve.
 The thick dashed line shows the theoretical prediction of Eq.~\eqref{eq:HOMsingleResult}
 for two infinitely narrow pulses at $T=$25mK.
 The dotted curve shows the shape of $V(t)$ over one period.}
\label{fig:HOMT}
\end{figure} 
 
Fig.~\ref{fig:HOMT} shows the normalized HOM noise for a periodic Lorentzian drive, at $\nu=1/3$,
 with realistic values for the experimental parameters (frequency $\omega = 1\times 2 \pi$ GHz, and finesse $\eta=0.01$).\footnote{Note that this frequency is a bit smaller than the one commonly used in experiments (for example $\omega \simeq 5 \times 2 \pi $ GHz in Ref.~\onlinecite{dubois13}). We have chosen realistic values for the experimental parameters,  which allows us to take a smaller value for the finesse, as higher harmonics of the base frequency $\omega$ are more easily accessed.
The experiment of Ref.~\onlinecite{dubois13}
used pulses with width as small as 30ps, which is similar to the width of 20ps that we consider
in Fig.~\ref{fig:HOMT}}
  The black dotted line shows the shape
of the narrow Lorentzian pulse over one period. 
The full curves show
the HOM dip as a function of the time-shift $\delta t$. In panel a), the average charge
per pulse is fixed to $q e =e/3$, and the temperature $T$ is varied from 250mK down to 25mK.
At $T = 25$mK, the hierarchy of the different time scales is thus: 
pulse width ($\sim 20 ps$) $\ll$ thermal time-scale ($\sim 300 ps$) 
 $<$ period ($\sim1000ps$).
One can readily see that, while the width of the HOM dip is close to that of the 
Lorentzian pulse at $T=$250mK, it significantly increases as the temperature is lowered,
ultimately being much larger at $T=$25mK. 
We consider, in panel b), a fixed temperature $T=$25mK,  
and an injected charge per period which varies from $q e=e$ down to $q e=e/3$. 
There,
the width of the HOM dip is similar to that of the incoming pulse for $q=1$
(corresponding to the injection of a full electron per period on each edge), before increasing substantially
as $q$ is lowered, recovering a wide HOM dip for $q=1/3$.
The thick dashed line corresponds to the analytical prediction of Eq.~\eqref{eq:HOMsingleResult}
for $T=25mK$.  This shows a very good agreement with the full numerical result obtained for $q=1/3$, with only a small underestimation of the width of the dip associated with the assumption of infinitely sharp pulses.

In conclusion, we have shown that the anyonic statistics of quasiparticles in the FQHE has direct
consequences on the HOM interference of excitations created by narrow voltage pulses.
Contrarily to the usual picture, where the width of the HOM dip is trivially related to the temporal extension of the incoming excitations, here it is fixed by the thermal scale, which dominates at low temperature. We have shown how this can be explained
by the anyonic braiding of the incoming quasiparticles with thermal excitations naturally occurring
at the QPC. 
Reducing temperature increases the thermal time, which enhances the time scale
on which braiding is effective, and thus leads to a wider HOM dip.
 Our proposal could be realized with current experimental techniques, and
could lead to an original and relatively simple way to observe directly the consequences of
anyonic statistics in the FQHE. A natural extension of this work would be to consider
more exotic fractions
like $\nu=2/5$ or $\nu=2/3$~\cite{Kane1994,bid10},  or even $\nu=5/2$, characterized by non-Abelian statistics.\cite{Lee2007,nayak08,dolev08,lee22}.

\acknowledgements We are grateful to G. F\`eve, C. Glattli, B. Pla\c{c}ais and U. Gennser for
enlightning discussions. This work was carried out in the framework of the project
``ANY-HALL'' (Grant ANR No ANR-21-CE30-0064-03). Centre de Calcul Intensif d’Aix-Marseille is acknowledged for granting access to its high performance computing resources for early parts of this work.
 
\bibliography{AnyonHOM_resub_short_condmat}{}

\vspace{5.cm}


\pagebreak

\onecolumngrid

\begin{center}
\begin{Large}
\pagebreak
\textbf{Supplemental material}
\end{Large}
\end{center}

\setcounter{equation}{0}
\setcounter{figure}{0}
\setcounter{table}{0}
\setcounter{page}{1}
\renewcommand{\theequation}{S\arabic{equation}}
\renewcommand{\thefigure}{S\arabic{figure}}
\renewcommand{\bibnumfmt}[1]{[S#1]}
\renewcommand{\citenumfont}[1]{S#1}

\section{Green functions and their properties}

The quasiparticle Green function is defined as
\begin{align}
{\cal{G}}_{R/L} \left (x,x' ; t^\eta, t'^{\eta'} \right) = \left\langle T_K \psi_{R/L}^\dagger \left( x,t^\eta \right) \psi_{R/L} \left( x',t'^{\eta'} \right)  \right\rangle .
\end{align}

Using the properties of time ordering, and the linear dispersion along the edge, this can be recast under the simplified form
\begin{align}
{\cal{G}}_{R/L} \left (x,x' ; t^\eta, t'^{\eta'} \right) =  {\cal{G}}_{R/L} \left(  \sigma^{\eta \eta'}_{t t'} \left( t - t' \mp \frac{x-x'}{v_F} \right) \right),
\end{align}
where $\sigma^{\eta \eta'}_{t t'}= \mbox{sign}(t-t') (\eta+\eta')/2 + (\eta'-\eta)/2$ and 
\begin{align}
{\cal{G}}_{R/L} (t) = \left\langle  \psi_{R/L}^\dagger (0,t) \psi_{R/L} (0,0)  \right\rangle .
\end{align}
Invoking the bosonization identity, this is further reduced as
\begin{align}
{\cal{G}}_{R/L} (t) &= \frac{1}{2 \pi a} \left\langle  e^{i \sqrt{\nu} \phi_{R/L}^\dagger (0,t)}   e^{-i\sqrt{\nu} \phi_{R/L}(0,0)}  \right\rangle = \frac{1}{2 \pi a} e^{\nu G_{R/L}(t)},
\end{align}
where we introduced the bosonic Green function $G_{R/L}(t) = \left\langle \phi_{R/L}^\dagger (0,t) \phi_{R/L} (0,0) \right\rangle$.

From the free Hamiltonian $H_0$, one can readily extract the corresponding Green function for the bosonic modes as
\begin{align}
G_{R/L} (t) = - \log \left[ \frac{\sinh \left( i \frac{\pi a}{\beta v_F} - \frac{\pi t }{\beta}\right)}{\sinh \left( i \frac{\pi a}{\beta v_F} \right)} \right] ,
\end{align}
so that the quasiparticle Green function ultimately reads
\begin{align}
{\cal{G}}_{R/L} (t) =  \frac{1}{2 \pi a} \left[ \frac{\sinh \left( i \frac{\pi a}{\beta v_F} \right)}{\sinh \left( i \frac{\pi a}{\beta v_F} - \frac{\pi t }{\beta}\right)}   \right]^\nu
\end{align}
One can easily show that this Green function is identical for right- and left-movers, so that we can safely drop the $R/L$ subscript from this point onward.

As anyons obey fractional statistics, they show nontrivial exchange properties which ensure that, at equal time, one has
\begin{align}
 \psi_R^\dagger (0,t) \psi_R(x,t) = e^{-i \pi \nu \text{Sign}(x)} \psi_R(x,t)   \psi_R^\dagger (0,t)
\end{align}
Making use of the linear dispersion along the edge, this is rewritten as
\begin{align}
 \psi_R^\dagger (0,t) \psi_R \left( 0,t-\frac{x}{v_F} \right) = e^{-i \pi \nu \text{Sign}(x)} \psi_R \left( 0,t-\frac{x}{v_F} \right)   \psi_R^\dagger (0,t)
\end{align}
Since this is valid for any set of parameters $(x,t)$, one can choose $x= v_F t$, without loss of generality. Taking then the quantum average, this yields
\begin{align}
\left\langle \psi_R^\dagger (0,t) \psi_R ( 0,0 ) \right \rangle &= e^{-i \pi \nu \text{Sign}(t)} \left\langle \psi_R(0,0)   \psi_R^\dagger (0,t) \right \rangle \nonumber \\
{\cal{G}} (t) &= e^{-i \pi \nu \text{Sign}(t)} {\cal{G}} (-t) 
\label{eq:ratioG}
\end{align}
It follows that the value of the ratio ${\cal{G}} (t)/{\cal{G}} (-t)$ can be viewed as a direct consequence of the exchange statistics of anyons.

\section{Computing the tunneling current}

\subsection{Tunneling current when injecting a single quasiparticle}

The tunneling current operator reads $I_T(t)= i e^* (\Gamma \psi_R^{\dagger}(0,t) \psi_L(0,t) - \mbox{H.c.})$. Here, we consider the situation where a single quasiparticle is incoming along the right edge, described by a prepared state of the form $| \varphi \rangle = \psi_R^\dagger (-x_0,-\mathcal{T}) | 0 \rangle$.

To lowest order in $\Gamma$, the mean current is thus given by
\begin{align}
\left \langle I_T(t) \right \rangle &= - \frac{i}{2} \int \!\! dt'
\sum_{\eta,\eta'}  \eta' 
\left\langle \varphi \left| T_K \;  I_T \left(  t^\eta \right) H_T \left( t'^{\eta'} \right) \right|\varphi \right\rangle \nonumber \\
 &= \frac{e^*}{2} \int \!\! dt'\sum_{\epsilon,\epsilon'}
\sum_{\eta,\eta'} \epsilon \eta' 
\langle 0 | T_K \; \psi_R (-x_0,-\mathcal{T^{-}})   \left( \Gamma \psi_R^{\dagger}(0,t^{\eta}) \psi_L(0,t^{\eta})\right)^ {(\epsilon)} 
  \nonumber\\
& 
\qquad \qquad \qquad \qquad  \times \left( \Gamma \psi_R^{\dagger}(0,t'^{\eta'})   \psi_L(0,t'^{\eta'})\right)^ {(\epsilon')} \psi_R^{\dagger}(-x_0,-\mathcal{T^{+}}) |0 \rangle
\end{align}
where $\epsilon=\pm$ is used to include the Hermitian conjugated terms, such that for $\epsilon=+$, one has for any operator $O$, $O^{(+)} = O$ while for $\epsilon=-$, one has $O^{(-)} = O^\dagger$.

Here, $T_K$ ensures the time-ordering along the Keldysh contour, and $\eta,\eta'=\pm$ are Keldysh indices. Note that we consider the injection of QP to have happened in the distant past. The Kelsdysh indices added to the times $-\mathcal{T}$ have been chosen to ensure that the $\psi_R (-x_0,-\mathcal{T}^{-})$ and $ \psi_R^{\dagger}(-x_0,-\mathcal{T}^{+}) $ operators remain in the same position after time ordering, independently of the values of $t$ and $t'$, for $\mathcal{T}$ large enough. In particular, keeping in mind that
$x_0 = v_F \mathcal{T}$ (corresponding to a quasiparticle reaching the QPC at $t=0$), this allows us to simplify some of the resulting Green functions as
\begin{align}
\mathcal{G} (-x_0, 0 ; -\mathcal{T}^-, t^\eta) &= \mathcal{G} (-t) \\
\mathcal{G} (0, -x_0 ; t^\eta, -\mathcal{T}^+) &= \mathcal{G} (t) ,
\end{align}
independently of $\eta$ and $t$, provided that $t \ll \mathcal{T}$.

Using the bosonized form of the quasiparticle operators, we have
\begin{align}
\left \langle I_T(t) \right \rangle = \Gamma^2 & \frac{e^*}{2} \int \!\! dt'\sum_{\epsilon}
\sum_{\eta,\eta'} \epsilon \eta'  \left[  \mathcal{G} \left(  \sigma_{tt'}^{\eta\eta'} \left(t-t'\right)  \right)  \right]^2  
  \left( \frac{\mathcal{G}(-t') \mathcal{G}(t)}{\mathcal{G}(t') \mathcal{G}(-t)}  \right)^{\epsilon} 
  \label{eq:ITQP}
\end{align}

Using the properties of the Green function derived in Eq.~\eqref{eq:ratioG}, this then becomes
\begin{align}
\left \langle I_T(t) \right \rangle = 2 i  e^* \Gamma^2 
\int_{-\infty}^{t} \!\! dt' \; 
\sin\left(2 \pi \nu \int_{t'}^{t} \!\!\! d\tau \; \delta(\tau)\right)
  \times \left[ \mathcal{G}(t-t')^2 - \mathcal{G}(t'-t)^2\right]  
\label{eq:Iqp_SM}
\end{align}
Changing the integration variable to $\tau=-t'$, and using the expression of the Green function, 
we get:
\begin{equation}
\left \langle I_T(t) \right \rangle = \theta(t) 2 i  e^* \frac{\Gamma^2}{(2 \pi a)^2}
  \sin(2 \pi \nu) 
  \int_0^{\infty} \!\!\!\! d\tau \left[
  \left(\frac{\mbox{sinh}(i \pi T \tau_0)}
{\mbox{sinh}(\pi T(i \tau_0 -t - \tau)}\right)^{2 \nu} - 
 \left(\frac{\mbox{sinh}(i \pi T \tau_0)}
{\mbox{sinh}(\pi T(i \tau_0 + t + \tau)}\right)^{2 \nu}
  \right] 
\end{equation}
where $\tau_0 = a/v_F$, $T = 1/(k_B \beta)$ is the temperature, and we use $k_B = \hbar =1$. Defining the reduced variables $\alpha = \pi T \tau_0$,
$u=\pi T \tau$ and $z = \pi T t$, the first term in the integral can be written as
\begin{align}
\int_0^{\infty} \!\!\! du  \left( \frac{\sinh(i \alpha)}{\sinh(i \alpha -z - u)}   \right)^{2 \nu}   &= \int_0^{\infty} \!\!\! du 
   \left( \frac{e^{i \alpha} - e^{-i \alpha}}{-e^{-i \alpha} e^{z}}  
   \frac{1}{1-e^{2 i \alpha}e^{-2 z} e^{-2 u}} \right)^{2 \nu} e^{-2 \nu \, u}  \\
   &= \left(e^{-z} \left(1-e^{2 i \alpha}\right) \right)^{2 \nu} 
       \int_0^{\infty} \!\!\! du \left( 1- e^{2 i \alpha} e^{-2 z} e^{-2 u} 
       \right)^{-2 \nu} e^{-2 \nu \, u} \\
   & =\frac{1}{2} \left(e^{-z} \left(1-e^{2 i \alpha}\right) \right)^{2 \nu}  
     \;  \left(\frac{1}{\nu}\right) \;  _2F_1\left(2 \nu,\nu,\nu+1,e^{2 i \alpha -2 z} \right)
\end{align}
where $_2F_1$ is the hypergeometric function. Using this result, the current can eventually be recast as
\begin{align}
\left \langle I_T(t) \right \rangle &= \theta(t) \; 2e^* 
 \left( \frac{\Gamma}{2 \pi v_F \tau_0} \right)^2
\frac{\sin(2 \pi \nu)}{2 \pi \nu T} \, e^{-2 \nu \pi T t} \,  \left( 2 \sin(\pi T \tau_0) \right)^{2 \nu}  \nonumber \\
 & \quad \quad \times 2 \mbox{Im} \left[ 
 _2F_1\left(2 \nu,\nu,\nu+1,e^{-2 \nu \pi T t} e^{-2 i \pi T \tau_0}\right)
 e^{i \pi \nu (1 - 2 T \tau_0)} \right]
\end{align}
Taking then the leading order in the cutoff parameter $\tau_0$ leads to
\begin{align}
\left \langle I_T(t) \right \rangle &= \theta(t) \; 2e^*  
 \left( \frac{\Gamma}{2 \pi v_F} \right)^2 \tau_0^{2 \nu -2} \; \frac{2 \sin(\pi \nu)  \sin(2 \pi \nu)}{\nu}
  \, e^{-2 \nu \pi T t}
 (2 \pi T)^{2 \nu -1} \; 
 \; _2F_1\left(2 \nu,\nu,\nu+1,e^{-2 \nu \pi T t}\right)
\end{align}
We see that this is a function of $2 \nu \pi T t=2 \nu \pi t/\beta$, which implies that the typical
length scale for this function is $\sim \beta$. The behavior of the current
in the two limits $t \ll \beta$ and $t \gg \beta$ is obtained by using the asymptotic
behavior of the hypergeometric function:
\begin{equation}
2F_1\left(2 \nu,\nu,\nu+1,e^{-2 \nu \pi T t}\right) = \left\{
\begin{array}{cc}
\nu \frac{\Gamma(\nu)^2}{\Gamma(2 \nu)} \frac{\sin(\pi \nu)}{\sin(2 \pi \nu)} -
\frac{\nu}{1-2\nu}\left( 1-e^{-2 \nu \pi T t} \right) &  \quad t \ll \beta \\
1+\frac{2\nu^2}{\nu+1}e^{-2 \nu \pi T t} & \quad t \gg \beta
\end{array}
\right.  .
\end{equation}

\subsection{Tunneling current when injecting a single electron}

It is instructive to repeat the same kind of derivation, only this time considering the situation where a single electron is incoming along the right edge. The prepared state now takes the form $| \varphi \rangle = \Psi_R^\dagger (-x_0,-\mathcal{T}) | 0 \rangle$, where the electron operator $\Psi_R$ satisfies the bosonization identity $\Psi_{R}(x) =\frac{ U_{R}}{2 \pi a}   e^{ i k_F x} e^{-i \frac{1}{\sqrt{\nu}} \phi_{R}(x)}$.

Following a similar derivation to the one above, one obtains instead of Eq.~\eqref{eq:ITQP}, the following expression for the tunneling current
\begin{align}
\left \langle I_T(t) \right \rangle = \Gamma^2 & \frac{e^*}{2} \int \!\! dt'
\sum_{\eta,\eta'} \epsilon \eta'  \left[  \mathcal{G} \left(  \sigma_{tt'}^{\eta\eta'} \left(t-t'\right)  \right)  \right]^2  
\left[
  \left( \frac{\mathcal{G}(-t') \mathcal{G}(t)}{\mathcal{G}(t') \mathcal{G}(-t)}  \right)^{1/\nu} 
  -
    \left( \frac{\mathcal{G}(t') \mathcal{G}(-t)}{\mathcal{G}(-t') \mathcal{G}(t)}  \right)^{1/\nu} 
\right]
\end{align}

From the properties of the quasiparticle Green function, Eq.~\eqref{eq:ratioG}, one readily sees that for $t\neq 0$
 \begin{equation}
  \left( \frac{\mathcal{G}(-t') \mathcal{G}(t)}{\mathcal{G}(t') \mathcal{G}(-t)}  \right)^{1/\nu}    = \mbox{exp}\left(-i \,  \int_{t'}^{t} \!\!\! d\tau \; 2 \pi \delta(\tau)\right) =1 ,
\end{equation} 
so that the tunneling current vanishes at all times $t \neq 0$ and is nonzero only at the specific time that the electron reaches the QPC.

\subsection{Tunneling current in the presence of a time-dependent voltage}

In the presence of a voltage bias, the tunneling part of the Hamiltonian can be written as
\begin{align}
H_T(t) & = \Gamma \, \mbox{exp}\left[i e^* \int_{-\infty}^{t} \!\!\!\!dt' \; V(t')\right]
   \psi_R^{\dagger}(0,t) \psi_L(0,t) + \mbox{H.c.}
\end{align}
where it now contains the effect of the applied votlage $V(t)$.

The tunneling current operator now reads
\begin{equation}
I_T(t)  = i e^*
 \left(\Gamma \, \mbox{exp}\left[i e^* \int_{-\infty}^{t} \!\!\!\!dt' \; V(t')\right]
   \psi_R^{\dagger}(0,t) \psi_L(0,t) - \mbox{H.c.}\right) .
\end{equation}
Taking the quantum average, the mean tunneling current is given in full generality by
\begin{align}
\left \langle I_T(t) \right \rangle &= 
\frac{i e^*}{2} \sum_{\eta} \sum_{\epsilon} \epsilon   
\bigg\langle T_{K}  \left( \Gamma \, \mbox{exp}\left[i e^* \int_{-\infty}^{t} \!\!\!\!dt' \; V(t')\right] \psi_R^{\dagger}(0,t^{\eta}) \psi_L(0,t^{\eta})  \right)^{(\epsilon)}   \nonumber \\ 
 & \quad \quad \quad \times \mbox{exp}\left[ -i \sum_{\eta'} \eta' \int_{-\infty}^{\infty} \!\!\!\! dt' \; H_T(t'^{\eta'}) \right] 
\bigg\rangle
\end{align}
where the sum on $\epsilon=\pm$ is used to represent the Hermitian conjugate,
and $\eta,\eta'=\pm$ are Keldysh indices.

Performing a perturbative expansion in the tunneling amplitude $\Gamma$, this gives up to second order
\begin{align}
\left \langle I_T(t) \right \rangle &= 
\frac{e^*}{2} \Gamma^2 \sum_{\eta,\eta'} \sum_{\epsilon} \epsilon \eta' 
 \int_{-\infty}^{\infty}\!\!\!\!dt' \;  \, 
 \mbox{exp}\left[i \, \epsilon \, e^* \int_{-\infty}^{t} \!\!\!\!dt' \; V(t')\right]
 \left \langle  T_K \psi_R^{\dagger}(0,t^{\eta})  \psi_R
(0,t'^{\eta'}) \right \rangle
   \left \langle  T_K  \psi_L(0,t^{\eta})  \psi_L^{\dagger}
(0,t'^{\eta'}) \right \rangle
\end{align}
Using the expression for the quasiparticle Green function, and performing explicitly the sum on the Keldysh indices $\eta$ and 
$\eta'$, one eventually gets
\begin{equation}
\left \langle I_T(t) \right \rangle =
2 i e^* \Gamma^2 
\int_{-\infty}^{t} \!\!\!\! dt' \; \sin \left(e^*\int_{t'}^{t} dt'' V(t'') \right) 
 \left[ \mathcal{G}(t-t')^2 - \mathcal{G}(t'-t)^2\right] .
\end{equation}
where the Keldysh summations end up restricting the $t'$ integral from $-\infty$ to $t$.

\section{Computing the noise}

\subsection{General expression}

The current noise is defined as:
\begin{equation}
S(t,t') =  \left \langle T_K \delta I_{T}(t^{-}) \, \delta I_{T}(t'^{+}) \right \rangle  
\label{eq:defnoise}
\end{equation}
with $\delta I_{T} (t) = I_{T}(t) - \left \langle I_{T} (t) \right \rangle$, and $\pm$ are
Keldysh indices.

In the presence of a voltage bias applied to both edges, the tunneling part of the Hamiltonian can be written as
\begin{align}
H_T(t) & = \Gamma \, \mbox{exp}\left[i e^* \int_{-\infty}^{t} \!\!\!\!dt' \; \left(V_R(t') - V_L(t') \right) \right]
   \psi_R^{\dagger}(0,t) \psi_L(0,t) + \mbox{H.c.}
\end{align}
where  we applied  a standard gauge transformation in order to reabsorb the effect of the voltage drives into the tunneling amplitude. In this situation, the tunneling current operator reads
\begin{equation}
I_T(t)  = i e^*
 \left(\Gamma \, \mbox{exp}\left[i e^* \int_{-\infty}^{t} \!\!\!\!dt' \; \left( V_R(t') - V_L(t') \right) \right]
   \psi_R^{\dagger}(0,t) \psi_L(0,t) - \mbox{H.c.}\right) .
\end{equation}
Substituting this back into Eq.~\eqref{eq:defnoise}, one readily obtains, up to lowest order in the tunneling amplitude $\Gamma$
\begin{equation}
S(t,t') =
2   \left ( \frac{e^* \Gamma}{2 \pi a} \right )^2 
 \cos \left( e^* \int_{t'}^{t} dt'' (V_R(t'') - V_L(t'') ) \right) 
 \mathcal{G}(t-t')^2.
 \label{eq:noisett}
\end{equation}
In what follows, we focus on the Hanbury-Brown Twiss (HBT) and the Hong-Ou-Mandel (HOM) setups, corresponding respectively to applying a single voltage drive, or to applying both of them.

\subsection{HOM noise for two narrow pulses of average charge $e^*$}

We consider here the case of two infinitely short pulses so that both $V_R(t)$ and $V_L(t)$ are composed of a single delta function,
with a time-shift $\delta t$ between them. Focusing on pulses of average charge $e^*$, one can thus write
\begin{equation}
V_R(t) = \frac{2\pi}{e} \delta \left( t+ \frac{\delta t}{2} \right) \quad \quad \quad V_L(t) = \frac{2\pi}{e} \delta \left( t- \frac{\delta t}{2} \right)  .
\end{equation}
The cosine factor entering the expression for the noise in Eq.~\eqref{eq:noisett} then simply reduces to either $\cos(2 \pi \nu)$ or to 1, depending on the values of $t$ and $t'$,  so that we write it as $\cos \left[ 2 \pi \nu f_{\delta t}(t,t') \right]$. The newly defined function $f_{\delta t}(t,t')$ is 1 if one of the times $t$ or $t'$ is in the interval $[-\delta t/2,\delta t/2]$ while the other one is not, and reduces to 0 otherwise.

The HOM noise is defined as the zero-frequency noise due to the collision of these two
excitations, as a function of the time-interval $\delta t$. Focusing on the zero-frequency contribution, and filtering out the equilibrium thermal noise (by subtracting the value in the absence of voltage drives), one has for the un-normalized HOM noise
\begin{equation}
\mathcal{S}_{HOM} = S (V_R,V_L) - S (0,0) = 2   \left ( \frac{e^* \Gamma}{2 \pi a} \right )^2  
  \int_{-\infty}^{\infty} \!\!\!\!dt \int_{-\infty}^{\infty} \!\!\!\!dt' 
  \left\{ \cos \left[ 2 \pi \nu f_{\delta t}(t,t') \right] -1 \right\} \mathcal{G}(t-t')^2
\end{equation}

Similarly, one can work out the expression for the corresponding noise when only one of the drives is present. The resulting HBT noise reads
\begin{equation}
\mathcal{S}_{HBT} = S (V_R,0) - S (0,0) = 2   \left ( \frac{e^* \Gamma}{2 \pi a} \right )^2  
  \int_{-\infty}^{\infty} \!\!\!\!dt \int_{-\infty}^{\infty} \!\!\!\!dt' 
 \left[ \cos \left( 2 \pi \nu \frac{1-\text{sign} (t\times t')}{2} \right) -1\right] \mathcal{G}(t-t')^2
\end{equation}

The standard HOM noise ratio is then defined as the ratio of the un-normalized HOM noise to twice the HBT noise, so that
\begin{align}
S_{HOM} (\delta t) &= \frac{\mathcal{S}_{HOM}}{2 \mathcal{S}_{HBT}} \nonumber \\
&= \frac{ \int dt dt' \left\{ \cos \left[ 2 \pi \nu f_{\delta t} (t,t') \right] -1\right\} e^{2 \nu G(t'-t)}}{2 \int dt dt' \left[ \cos \left( 2 \pi \nu \frac{1-\text{sign} (t\times t')}{2} \right) -1\right] e^{2 \nu G(t'-t)}}
\end{align}

Substituting the actual value of $f_{\delta t} (t,t') $, this can be further rewritten as
\begin{align}
S_{HOM} (\delta t) &= \frac{ \int_0^{\left| \delta t \right|} dt \int_0^{\infty} dt' \left[ e^{2 \nu G(t + t')} + e^{2 \nu G(-t-t')} \right]}{ \int_0^{\infty} dt \int_0^{\infty} dt' \left[ e^{2 \nu G(t + t')} + e^{2 \nu G(-t-t')} \right]} \nonumber \\
&= \frac{\text{Re} \left[   \int_0^{\left| \delta t \right|} dt \int_0^{\infty} dt'  e^{2 \nu G(t + t')}  \right]}{\text{Re} \left[   \int_0^{\infty} dt \int_0^{\infty} dt'  e^{2 \nu G(t + t')}  \right]} \nonumber \\
&= 1 - \frac{\text{Re} \left[  {\cal I} \left( \delta \right) \right]}{\text{Re} \left[  {\cal I} \left( 0 \right) \right]}
\label{eq:SHOMint}
\end{align}
where we introduced
\begin{align}
{\cal I} \left( \delta \right) &= \int_0^\infty dz ~z \left( \frac{\sinh (i \alpha)}{\sinh(i \alpha - z - \delta)} \right)^{2 \nu}
\end{align}
with the reduced variable $\delta = \pi \left| \delta t \right|/ \beta$, and the infinitesimal $\alpha = \pi \tau_0/\beta$.

This integral can be worked out as
\begin{align}
{\cal I} \left( \delta \right) &= -\frac{1}{4} \left( 1 - e^{2 i \alpha} \right)^{2 \nu} 
e^{-2 \nu \delta} \partial_\gamma \left[ \frac{1}{\nu+\gamma}  
{_2}F_1
\left( 2 \nu, \nu+\gamma ; \nu+\gamma+1 ; e^{2 i \alpha} e^{-2 \delta}\right)
 \right]_{\gamma = 0}
\end{align}
where one clearly sees that for $\delta \ll 1$, the exponential prefactor dominates, so that
\begin{align}
{\cal I} \left( \delta \right) &\underset{\delta \ll 1}{\simeq}   e^{-2 \nu \delta} {\cal I} \left( 0 \right)
\end{align}
It follows that, in the regime where $| \delta t|/\beta \to 0$, one has
\begin{align}
S_{HOM} (\delta t)  &\underset{| \delta t|/\beta \to 0}{\longrightarrow} 1-  e^{-2 \pi \nu  \frac{| \delta t |}{\beta}} 
\end{align}

\subsection{HOM noise for two narrow pulses of  average charge $q e$}

The previous results can be easily extended to the case of pulses carrying a different charge. We now define
\begin{equation}
V_R(t) = \frac{2\pi q}{\nu e} \delta \left( t+ \frac{\delta t}{2} \right) \quad \quad \quad V_L(t) = \frac{2\pi q}{\nu e} \delta \left( t- \frac{\delta t}{2} \right)  .
\end{equation}

Following the lines of the previous calculation, one can similarly obtain an expression for the HOM noise ratio as
\begin{align}
S_{HOM} (\delta t) 
&= \frac{ \int dt dt' \left\{ \cos \left[ 2 \pi q f_{\delta t} (t,t') \right] -1\right\} e^{2 \nu G(t'-t)}}{2 \int dt dt' \left[ \cos \left( 2 \pi q \frac{1-\text{sign} (t\times t')}{2} \right) -1\right] e^{2 \nu G(t'-t)}}
\end{align}
Interestingly, while the resulting integrals are finite for different domains in time, they always contain a prefactor $\cos (2 \pi q)-1$. For $q \notin \mathbb{Z}$, this prefactor simplifies between numerator and denominator, leaving us with the same expression as Eq.~\eqref{eq:SHOMint}, independently of $q$. This, however, is specific to the very short pulse situation, as a finite extent leads to slightly different contributions for the numerator and denominator, which depend on $q$ in a nontrivial way.


\subsection{HOM noise in the Floquet formalism}

The applied voltages on the right and left edges are now given by periodic Lorentzian pulses. They are identical except for a time-shift $\delta t$, so that
  \begin{equation}
   V_L(t) = V_R(t-\delta t)= \frac{V_{DC}}{\pi} \sum_k \frac{\eta}{\eta^2+(t/T_0-k)^2}
  \end{equation} 
  In  the Floquet formalism, the essential ingredients are the coefficients $p_l$, which 
 are the Fourier components of the accumulated phase $\phi(t) = e^* \int_{-\infty}^t dt' V_{AC}(t')$
 created by the AC part of the time-dependent voltage. In practice, it is convenient to introduce the  time-dependent voltage  $V_\text{diff}(t)=V_R(t) - V_L(t)$ which naturally appears in the expression of the noise.
 
 Starting back from the general expression of Eq.~\eqref{eq:noisett}, and inserting the $p_l$ coefficients associated with a generic drive $V(t)$ (this allows us to replace $V$ with $V_R$, $V_L$ or $V_\text{diff}$), one can write
 \begin{align}
S(t,t') &= 2   \left ( \frac{e^* \Gamma}{2 \pi a} \right )^2 
 \cos \left[ e^* \int_{t'}^{t} dt'' V (t'')  \right] 
 \mathcal{G}(t-t')^2 \nonumber \\
 &=   \left ( \frac{e^* \Gamma}{2 \pi a} \right )^2 \sum_{l,m} p_l^* p_m
 \left( e^{i e^* V_{DC} (t-t')} e^{i l \omega t}  e^{-i m \omega t'}  + e^{-i e^* V_{DC} (t-t')}  e^{-i m \omega t}   e^{i l \omega t'}  \right)
 \mathcal{G}(t-t')^2
 \end{align}
 where $\omega = \frac{2 \pi}{T_0}$ is the frequency of the drive.
 
 In this Floquet formalism, the zero-frequency noise is now defined as
 \begin{align}
 \mathcal{S} = \int d\tau \int_0^{T_0} \frac{d \bar{t}}{T_0} S \left( \bar{t} + \frac{\tau}{2}, \bar{t} - \frac{\tau}{2} \right)
 \end{align}
 which becomes
 \begin{align}
 \mathcal{S}& = \int d\tau \int_0^{T_0} \frac{d \bar{t}}{T_0} 
 \left ( \frac{e^* \Gamma}{2 \pi a} \right )^2 \sum_{l,m} p_l^* p_m
 \left( e^{i e^* V_{DC} \tau} e^{i l \omega \left( \bar{t} + \frac{\tau}{2}\right)}  e^{-i m \omega \left( \bar{t} - \frac{\tau}{2} \right)}  + e^{-i e^* V_{DC} \tau}  e^{-i m \omega \left( \bar{t} + \frac{\tau}{2} \right)}   e^{i l \omega \left( \bar{t} - \frac{\tau}{2} \right)}  \right)
 \mathcal{G}(\tau)^2 \nonumber \\
 & = 2
 \left ( \frac{e^* \Gamma}{2 \pi a} \right )^2 \sum_l \left| p_l \right|^2  
 \int d\tau  \cos \left[ (l+q) \omega  \tau  \right]
 \mathcal{G}(\tau)^2 
 \end{align}
 where we introduced the average charge $q = \frac{e^* V_{DC}}{\omega}$ injected by the drive over one period.
 
 Introducing the coefficients  $p_{\text{diff},l}$ for the voltage
 difference $V_\text{diff}(t)$, as well as the coefficients $p_{L,l}$ and $p_{R,l}$ 
 corresponding to $V_L(t)$ and $V_R(t)$ applied individually, and noticing that $V_{R,DC} = V_{L, DC} = \frac{q \omega}{e^*}$, while $V_{\text{diff},DC}=0$, one finally has for the HOM noise ratio
 \begin{align}
 S_{HOM}(\delta t) = 
 \frac{\mathcal{S}_{HOM}}{2 \mathcal{S}_{HBT}} 
 &= \frac{\sum_l   F (p_{\text{diff},l} , 0) - \left| \Gamma(\nu) \right|^2}
  {\sum_l \left[ F(p_{L,l}, q) + F(p_{R,l}, q) \right] - 2  \left| \Gamma(\nu) \right|^2 }
  \label{eq:HOMpl} 
  \end{align}
  with
\begin{align}  
  F(p_l,q) & =  |p_l|^2 \left| \Gamma \left( \nu + i \frac{l+q}{2 \pi \theta} \right) \right|^2
    \mbox{cosh} \left(\frac{l+q}{2 \theta} \right)
 \end{align}
and $\theta = k_B T/\hbar \omega$ is the reduced temperature. Note that this expression is very general and can describe any kind of periodic potentials, provided that one uses the correct corresponding expressions of the $p_l$ coefficients.

\end{document}